\documentclass[11pt,preprint,graphicx]{aastex}
\def\etal{\it et al. \rm }
\begin{document}

\title{Tests of Chemical Enrichment Scenarios in Ellipticals Using
Continuum Colors and Spectroscopy}

\author{James Schombert}
\affil{Department of Physics, University of Oregon, Eugene, OR 97403;
js@abyss.uoregon.edu}

\author{Karl Rakos}
\affil{Institute for Astronomy, University of Vienna, A-1180, Wien, Austria;
karl.rakos@chello.at}

\begin{abstract}

We combine spectroscopic metallicity values with integrated narrowband
continuum colors to explore the internal metallicity distribution in
early-type galaxies.  The different techniques for determining
metallicity (indices versus colors) allows for an estimate of the
contribution from metal-poor stars in a predominantly metal-rich population
which, in turn, places constraints on the shape and width of a galaxy's
metallicity distribution function (MDF).  The color-spectroscopic data is
compared to the closed box, infall and inhomogeneous chemical evolution
models.  The G-dwarf problem, a deficiency in metal-poor stars as compared
to closed box models, is evident in the dataset and indicates this deficiency is
common to all early-type galaxies.  However, even simple infall models
predict galaxy colors which are too blue compared to the observations.  A
simple analytic model is proposed which matches the elliptical data and
recent HST observations of M31 (Worthey \etal 2005) and NGC 5128
(Harris \& Harris 2000) by reducing the number of metal-poor stars in a
systematic fashion.  While without physical justification, the shape of
these models are similar to predictions of inhomogeneous enrichment
scenarios.

\end{abstract}


\section{INTRODUCTION}

The chemical history of galaxies has been a key astrophysics issue since
the discovery that our own Galaxy is separated into kinematically distinct
regions by metallicity (Baade 1944, Gilmore, Wyse \& Kuijken 1989).  Due to
the process of enrichment by SN and AGB mass-loss, plus a star formation
rate that is extended in time, the metallicity distribution function (MDF)
in a galaxy evolves with time such that the canonical view of galaxy
evolution is that the enrichment process continues to increasingly higher
yields until the onset of a galactic wind (powered by SN input) halts star
formation (Matteucci 2007).  Thus, the MDF of stars becomes a map of past
galaxy evolutionary mechanisms.

For nearby galaxies, a direct examination of the color-magnitude diagram
(CMD) provides the clearest view of chemical evolution (Dolphin 2002,
Skillman \etal 2003, Harris \& Zaritsky 2004, Grebel 2004).  Fortunately,
the section of the CMD most sensitive to metallicity effects is the RGB,
which is brightest part of a galaxy's CMD and visible to the largest
distances.  This situation differs from methods to determine a galaxy's age
which are dependent on measurements of the fainter turnoff portion of the
CMD.

For more distant galaxies, we are forced to deduce the metallicity of the
underlying stellar population by comparison of colors or spectral indices
interpreted by spectroscopic evolutionary distribution (SED) models.
Unfortunately, this type of analysis produces only a luminosity weighted,
mean metallicity, although there can be some spatial information (i.e.
color gradients).  Color information does not resolve the shape of the
metallicity distribution, which is a critical piece of information in order
to test the type of chemical evolution that a galaxy has undergone.

In the last decade, a series of new techniques to investigate the
metallicity of stellar populations in galaxies allows for the possibility
of extracting some limited information on the shape of the MDF in
ellipticals.  It has been noted by several studies that the colors of
ellipticals are extremely uniform (Smolcic \etal 2006), yet can not be
matched with a SED model that is composed of a single metallicity stellar
population (Rakos, Schombert \& Odell 2008).  In addition, their colors, as
compared to SED models, indicate a contribution from a bluer (i.e.
possibly metal-poor) component (Worthey, Dorman \& Jones 1996, Rakos \&
Schombert 2008), in line with expectations from the MDF of stars in nearby
galaxies.

To resolve the effect of a metal-poor population on galaxy colors requires
a model of the MDF plus SED spectra of each metallicity bin to sum into an
integrated color.  In addition, an independent measure of metallicity is
required to distinguish the observed value of the colors from the colors
present by a population of singular metallicity.  Fortunately, just such a
measurement of [Fe/H] exists from spectroscopic studies, the $<$Fe$>$
index.  This index has the advantage of skipping the need for SED models
(although an MDF will produce a luminosity averaged value of $<$Fe$>$) and
directly measures the abundance of Fe.  Since Fe is the primary contributer
to line blanketing on the RGB, it is also the primary link to color changes
in galaxies from metallicity effects.

The goal of this project is to take advantage of the different manner in
which metallicity is determined in galaxies (i.e. colors versus
spectroscopy) to deduce the basic shape of the MDF in early-type galaxies.
To achieve this goal, we will require a series of predicted MDF's, given by
various chemical evolutionary scenarios, combined with SED models to
calculate the expected galaxy colors.  Thus, we will first examine the
quality of the current generation of SED models with respect to globular
clusters colors and metallicity for a set of special narrowband and near-IR
filters (a modified Str\"omgren system, Rakos \& Schombert 1995).  Second,
we will present a subset of galaxies from spectroscopic studies where we
have matching narrowband colors, and explore the behavior of color versus
$<$Fe$>$ for this sample.  Lastly, we will compare the resulting [M/H]
versus color plane with respect to the predictions of various chemical
evolution scenarios (such as closed box, infall, inhomogeneous) in order to
determine which scenario most closely matches the data.

\section{Analysis of SED Models}

In order to interpret the global colors or line indices in galaxies, one
needs 1) a star formation history model which includes not only a star
formation rate but a chemical enrichment model so that a present-day galaxy
is characterised by the sum of light from stars with a range of age and
metallicity plus 2) accurate SED's for each stellar population of a
particular age and metallicity.  Clearly, we wish to deduce the star
formation history of a galaxy by first determining the present-day mixture
of internal stellar populations parameterized by their mean age and mean
metallicities.  These stellar populations have unique luminosities by mass
which can be summed to determine the total color (or luminosity weighted
line indice) for the galaxy.

The first step in unraveling the underlying populations in galaxies is to
test the predictions of the various SED models in the literature against
systems where we have detailed knowledge of the age and metallicity of the
stars (i.e. stellar clusters).  The SED models take on the simplest
assumptions, a single burst resulting in a population with a single value
for their age and metallicity ($[M/H]$), a so-called single stellar
populations (SSP).  Their direct comparison to galaxy colors is problematic
as galaxies cannot be composed of a single burst population simply due to
the physics of initial galaxy formation as demonstrated by the discovery
of color/metallicity gradients across all Hubble types (see \S2.2).

There have been numerous SED models published in the literature since the
earliest attempts to model galaxies (Faber 1972, O'Connell 1976, Pickles
1985).  Our own narrowband photometry work (Rakos, Schombert \& Odell 2008)
has focused on the Bruzual \& Charlot (2003, hereafter BC03) models and the
models from the Gottingen group (Schulz \etal 2002).  The choice of these
two groups was for convenience as we used the original Bruzual \& Charlot
models for our study of distant clusters (Rakos \& Schombert 1995) and we
later switched to the Gottingen models since they published Str\"omgren
colors ($uvby$) that were easy to transform into our modified Str\"omgren
system ($uz$,$vz$,$bz$,$yz$).  The adoption of the latest stellar tracks
from the Padova group (Girardi \etal 2000) by both BC03 and Schulz has
rendered their models to be nearly identical with only minor differences in
resulting colors and line indices.  Thus, for the remainder of this paper
we will use the model tracks from BC03 for our analysis and only note the
differences between the BC03 and Schulz models where relevant.

For this work, we have used we have selected a fairly standard range of
models with an age of 12 Gyrs and [Fe/H] from $-$2.3 to $+$0.4, all using
the Chabrier (2003) IMF (mass cutoff at 0.1 and 100 $M_{\sun}$) and Padova
(2000) isochrones.  Each SSP is interpolated at the 0.1 dex level in
metallicity and convolved to our narrow band filters to produce a full grid
of colors.  These SSP's are then convolved into our various metallicity
distribution models as will be discussed in a later section.

\subsection{Comparison of SSP Models to Globular Clusters}

There do exist stellar systems which closely resemble the narrow conditions
of an SSP, the globular clusters of our own Galaxy (hereafter GC's).  Their
limited range of internal metallicity and age, as determined from detail
studies of their color-magnitude diagrams (CMD), indicates a constant
metallicity to within $\Delta{[Fe/H]} <$ 0.2 and spread in internal age of
less than 1 Gyr (Bruzual 2002).

A first test to the accuracy of an SED model is to match to the isochrones
of globular clusters, for which the modern formulations in the literature are
all adequate (see the analysis in Schiavon 2007).  Second, is to match the
integrated colors of SSP's to colors of globular clusters.  With respect to
the BC03 models, there is an excellent discussion of the model fits, as
compared to broadband colors, in Bruzual (2002).

As our work deals with integrated narrowband colors, we have collected a
sample of 32 GC's with high quality narrowband colors ($uz$,$vz$,$bz$,$yz$,
Rakos \& Schombert 2005) and combined this sample with a subset of GC's
with near-IR colors from Cohen \etal (2007).  The resulting six color-color
diagrams are shown in Figure 1 along with the SSP models of BC03 for an age
of 12 Gyrs.  The change in color is due solely to changes in metallicity
([Fe/H]) as the range of age for GC's is less than 2 Gyrs (Salaris \& Weiss
1998) and there is a negligible change in the SSP models for this age
range.  Note that the $uz$,$vz$,$bz$,$yz$ differs from the normal
Str\"omgren ($uvby$) system in the sense that the filters are slightly
narrower (by 20\AA) and the $uz$ filter is shifted 30\AA\ to the red in its
central wavelength as compared to the original system.  The $uz,vz,bz,yz$
system covers three regions in the near-UV and blue portion of the
spectrum.  The first region is longward of 4600\AA, where in the influence
of absorption lines is small.  This is characteristic of the $bz$ and $yz$
filters ($\lambda_{eff}$ = 4675\AA\ and 5500\AA).  The second region is a
band shortward of 4600\AA, but above the Balmer discontinuity.  This region
is strongly influenced by metal absorption lines (i.e. Fe, CN) particularly
for spectral classes F to M, which dominate the contribution of light in
old stellar populations.  This region is exploited by the $vz$ filter
($\lambda_{eff} = 4100$\AA).  The third region is a band shortward of the
Balmer discontinuity or below the effective limit of crowding of the Balmer
absorption lines.  This region is explored by the $uz$ filter
($\lambda_{eff} = 3500$\AA).

\begin{figure}
%\plotfiddle{f1.eps}{0.0truecm}{0}{400}{550}{25}{0}
\centering
\includegraphics[scale=0.95]{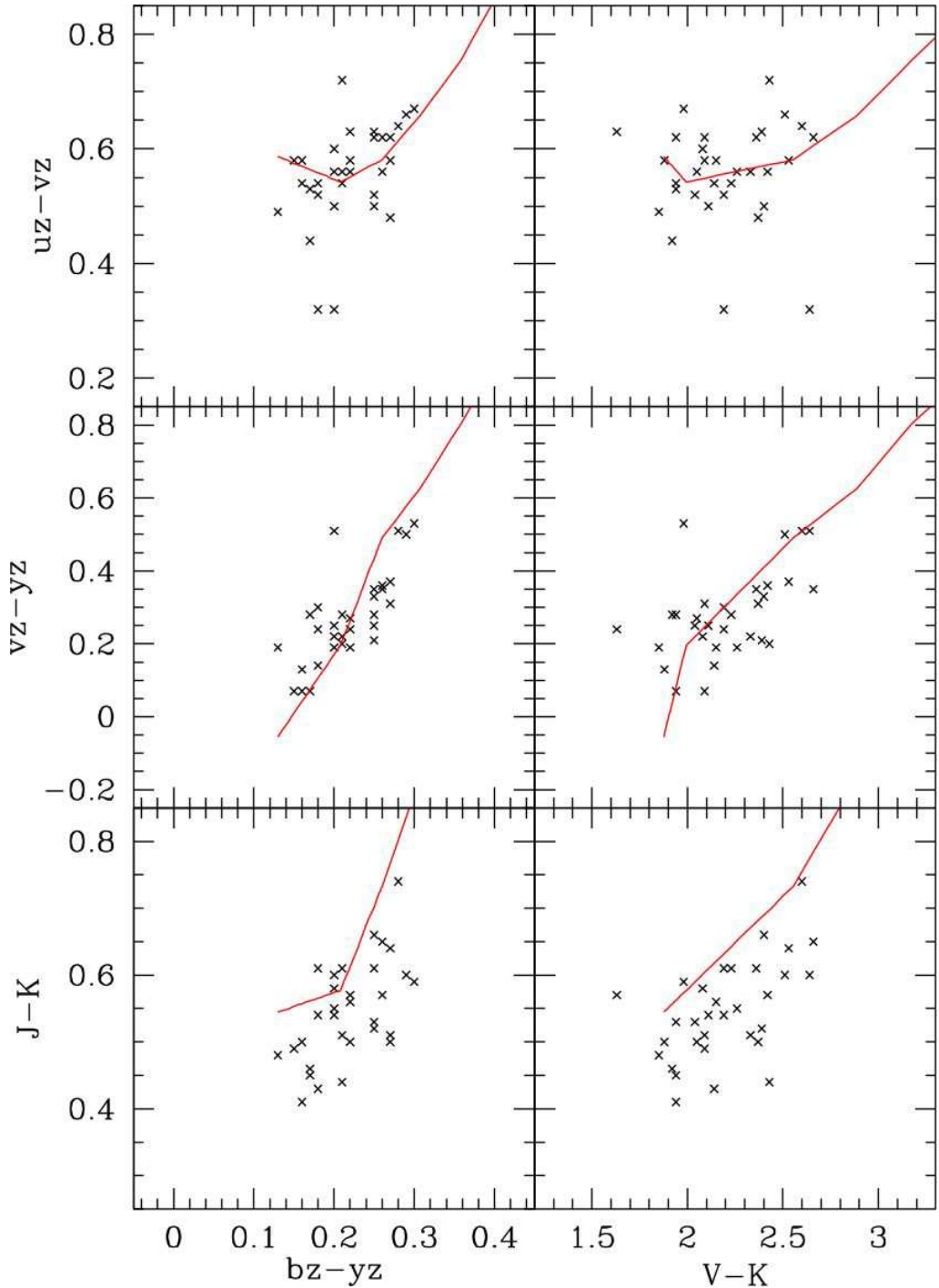}
\caption{The optical and near-IR color-color diagrams for Galactic globular
clusters (Rakos \& Schombert 2005).  As these clusters are nearly coeval in
age, the correlations in color reflect metallicity variations between the
clusters.  Also shown are the 12 Gyr SSP models from Bruzual \& Charlot
(2003).  The models are an excellent match to the data except for the
$J-K$ colors (for unknown reasons).  Scatter around the models is fully
explained by stochastic variation in color due to the number of stars at
the tip of the RGB and AGB.
}
\end{figure}

The scatter in optical and near-IR colors is relatively large considering
the photometric accuracy of the observations.  However, this is not a
statement concerning the quality of the models but rather an observation
effect due to the stochastic aspect of stellar light from GC's.  While a GC
may be composed of over $10^5$ stars, the stars at the tip of the
asymptotic giant branch (AGB) and horizontal branch (HB) dominate the
integrated light and, therefore, small variations in the number of these
stars can dramatically alter the measured colors (see an excellent
discussion of this effect in Bruzual \& Charlot 2003).  Given this
stochastic scatter, the models display a good match to the GC colors.  The
only serious discrepancy is found in the $J-K$ color, for reasons that are
not immediately apparent.  The clearest correlation in color is found
between our two optical colors ($vz-yz$, $bz-yz$) and the near-IR $V-K$
color.  These are the three colors most sensitive to metallicity changes
and it is not surprising to find them strongly correlated.

The correlations between color and metallicity are shown in Figure 2, plots
of color versus [Fe/H].  For consistency, we have adopted a calibration
from the total metallicity ([M/H]) from [Fe/H] based on the Padova 2000
tracks (Girardi \etal 2000, discussed in Bruzual \& Charlot (2003).  From
these plots, it is clear that colors in the near-UV region of the spectrum
are degenerate with respect to metallicity.  Near-IR colors are well fit by
the models, but have shallow slopes at low metallicities making their use
problematic.  The steepest slopes are seen in $vz-yz$ and $bz-yz$, with the
tightest correlation found for $vz-yz$.  This confirms what we have
learned in our earlier studies on the uniformity in color for ellipticals and the
correlations with galaxy mass (i.e. the color-magnitude relation, Odell,
Schombert \& Rakos 2002).  As our previous work on the age and metallicity
of ellipticals has demonstrated (Rakos, Schombert \& Odell 2008), the range
of age in ellipticals is very limited and the majority have mean ages
greater than 10 Gyrs.  For these ages, $vz-yz$ is strictly a measure of
metallicity and, thus, becomes our primary tool for determining metallicity
in SSP systems (such as GC's) and composite systems (such as S0's and
ellipticals).  This will be discussed further in \S3.

\begin{figure}
%\plotfiddle{f2.eps}{0.0truecm}{0}{400}{550}{25}{0}
\centering
\includegraphics[scale=0.95]{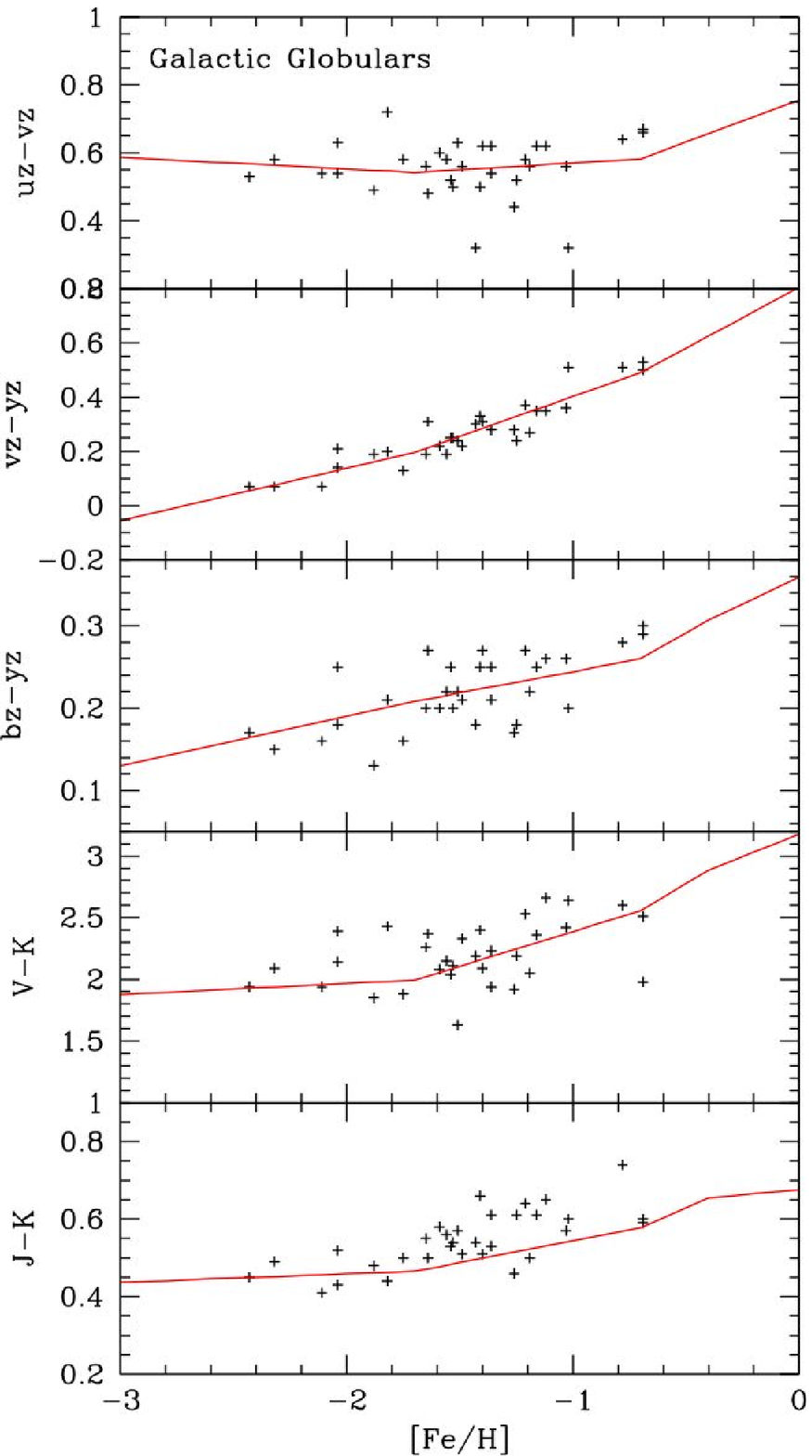}
\caption{Optical and near-IR colors versus metallicity ([Fe/H]) for 32
Galactic globulars.  The tightest correlation between color and [Fe/H] is
found for the narrowband color, $vz-yz$.  The $uz-vz$ color is influence by
hot HB stars and the second parameter effect.  The $bz-yz$ color lacks
sufficient metal lines for a stronger correlation.  Also shown are the 12
Gyr SSP models.
}
\end{figure}

\subsection{Composite Stellar Populations}

Galaxies are known to be composed of more than a simple stellar population
(i.e. single age and metallicity).  This is clearly the case for the Milky
Way, based on studies of nearby stars (Twarog 1980).  And metallicity
gradients in ellipticals demonstrate that they too are composed of stars
with a range of metallicities (Sanchez-Blazquez \etal 2006).  With respect
to integrated optical and near-IR colors, ellipticals have colors that are
clear extrapolations from SSP colors, such as galactic globulars, but have
significant differences that demonstrate that they are composed of stars
with a range of metallicities.  For example, the mid-ultraviolet region of
an elliptical's spectra are best modeled by an old metal-rich plus smaller
old metal-poor populations (Bressan, Chiosi \& Fagotto 1994, Rose \& Deng
1999, Lotz, Ferguson \& Bohlin 2000).

This composite color effect is best seen in Figure 3, a plot of our optical
$vz-yz$ and $bz-yz$ colors and the near-IR color $V-K$.  The globular
cluster data from Figure 2 is replotted along with a BC03 12 Gyr SSP
models, a good match to the GC data.  Also shown are the colors of 50
bright ellipticals in the core of the Coma cluster.  The optical colors for
the Coma sample derive from Odell, Schombert \& Rakos (2002), the near-IR
colors for the same galaxies are taken from Eisenhardt \etal (2007).  While
the SSP models are a good fit to the GC data, they do not fit the
elliptical colors.  The deviations from an SSP model are such that
ellipticals have $vz-yz$ colors that are slightly bluer than single
metallicity models (Rakos, Schombert \& Odell 2008).  This is exactly what
one would expect by ignoring the contribution of low metallicity stars and
a simple model that considers a range of metallicities is shown as a dashed
line in Figure 3.  A slightly younger SSP (7 Gyrs in Figure 3), would match
the optical colors, but fail to match the near-IR colors.  While this
simple composite model will be discussed in greater detail below, it
demonstrates that elliptical colors are best explained by a composite of
underlying metallicities, an obvious conclusion based on
population studies of our own Galaxy.

\begin{figure}
\centering
\includegraphics[scale=0.70,angle=-90]{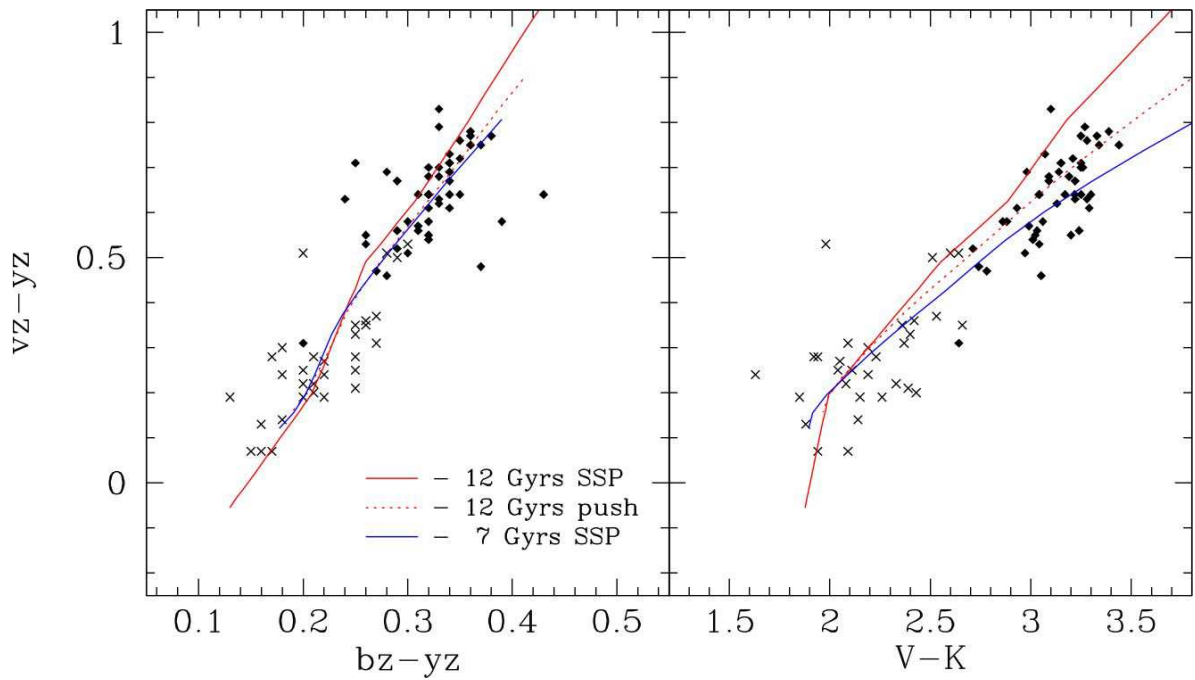}
\caption{Optical and near-IR color-color diagrams for globular clusters
(crosses) and Coma ellipticals (solid symbols).  The solid red line is a 12
Gyr SSP model, the blue line is a 7 Gyr SSP model.  While the GC data is
adequately described by the SSP models, a majority of ellipticals lie to
the right of the 12 Gyr model.  The dashed line is a composite stellar
population (CSP) model, for a 12 Gyr population using a simple infall
chemical enrichment scenario.  While a younger model can explain the
optical colors, a composite population is a better match to both the
optical and near-IR elliptical data.  While the colors of ellipticals are
extensions of globular cluster colors, these slight differences 
are best explained by a CSP model.
}
\end{figure}

Decoding the color of galaxies requires a map of the ages of the stellar
populations and their metallicity distribution (as a function of age).  For
the purposes of this initial examination of a galaxy's underlying
metallicity distribution, we will assume that all the color change in their
integrated luminosity is due solely to metallicity effects.  There are many
reasons to believe that a majority of the stars in ellipticals are old
($\tau$ $>$ 10 Gyrs) ranging from the tight correlation of the
color-magnitude relation to the red envelope in high redshift studies of
clusters.  In a parallel paper (Schombert \& Rakos 2009), we present a
detailed analysis of spectroscopic determination of galaxy ages as it
impacts on the observed color properties of ellipticals, i.e. the
color-magnitude relation (CMR).  Spectroscopic studies find a high fraction
of ellipticals with ages less than 7 Gyrs (see review by Schiavon 2007);
however, these ages are in conflict with the colors of early-type galaxies.
Detailed comparison to the CMR demonstrates that early-type galaxies with
ages less than 7 Gyrs are rare and we have ignored galaxy age from our
analysis.

\subsection{Chemical Evolution Models}

In order to model galaxy integrated colors, we will need to combine the
predicted colors from the SED models with a model of the internal
metallicity distribution, the so-called metallicity distribution function
(MDF, Pagel 1997).  The mean metallicity, $Z$, of a galaxy will then be a
luminosity weighted sum of the contribution from a continuum of
metallicities.  And it is expected that the peak $Z$ value will vary with
position in the galaxy (i.e. gradients) and with the mass of the galaxy,
such that higher mass galaxies process more material before the onset of
galactic winds overcomes their gravitational potentials to halt star
formation and further enrichment.

There are only a few galaxies with actual MDF's measured from HST imaging
of the tip of the RGB.  The most relevant examples to this study are the
old populations in M31 (Worthey \etal 2005) and the nearby elliptical NGC
5128 (Harris \& Harris 2000).  Both studies display MDF's with several
features in common; 1) a well defined gaussian-like peak, 2) a long tail to
low metallicities, 3) a sharp cutoff on the high metallicity side.  Both
galaxies display a lower metallicity peak with increasing radius from the
galaxy center (i.e. metallicity gradients).  This results in a narrower MDF
at lower peak metallicities.

The simplest model of chemical evolution to produce a MDF is, of course,
the closed-box enrichment scenario (van den Bergh 1962).  This scenario
assumes no infall or outflow of gas and the metallicity of the stars
increases in yield with every epoch of star formation.  The solution for
this model is analytic and displayed in Figure 4.

\begin{figure}
%\plotfiddle{f4.eps}{0.0truecm}{0}{400}{400}{25}{0}
\centering
\includegraphics[scale=0.95]{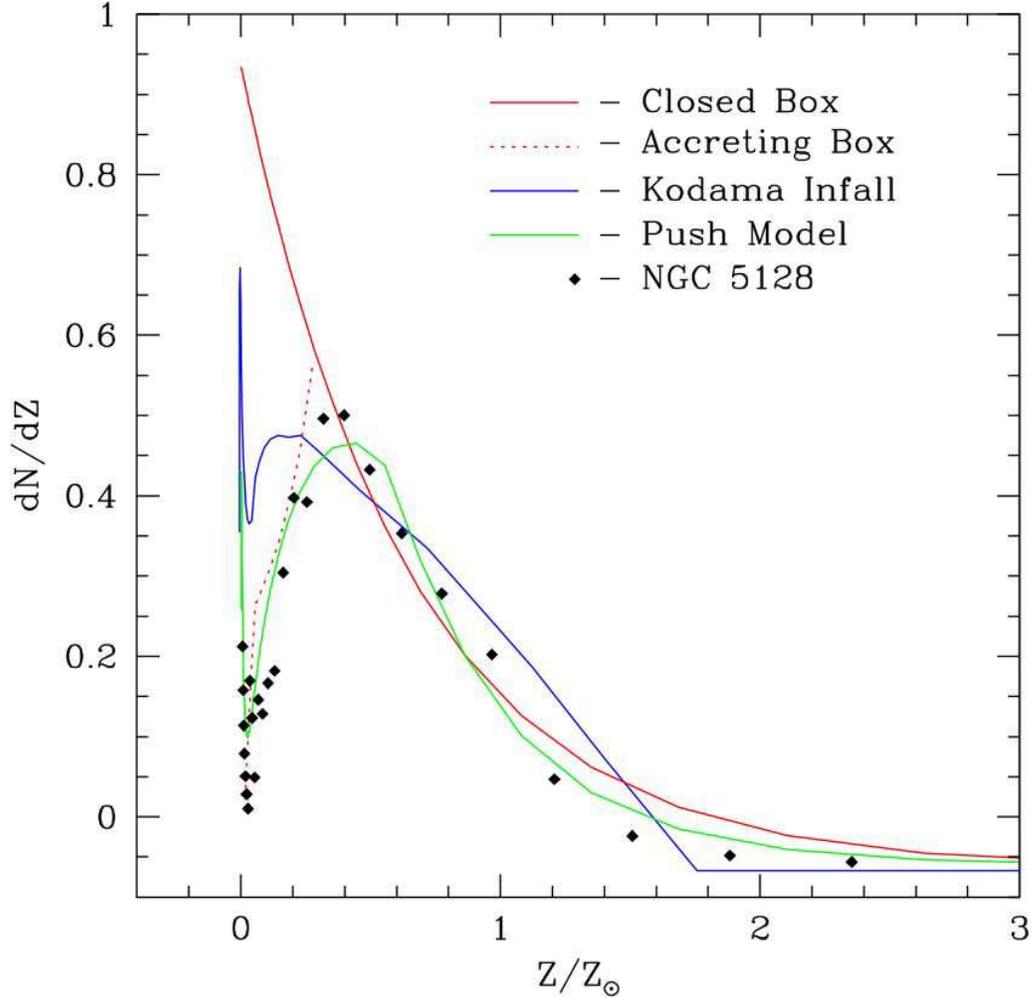}
\caption{Comparison of various chemical enrichment models.  The closed box
model is shown as a solid red line (with an initial enrichment as a red
dashed line).  An infall model (Kodama \& Arimoto 1997) is shown as a solid
blue line (see Yi \etal 1998).  Our analytic `push' model, designed to
artificially reduce the metal-poor component, is shown as a solid green
line.  All these models are compared to the metallicity data for the
elliptical NGC 5128 (inner regions, Harris \& Harris 2000).
}
\end{figure}

A well know problem for closed-box model is that it over estimates the
number of low metallicity stars ($Z/Z_{\sun} < 0.3$), the so-called G dwarf
problem (Gibson \& Matteucci 1997).  This over estimation occurs not only
for stars in the local solar neighborhood (van den Bergh 1962, Schmidt
1963) but also in the MDF's in nearby galaxies (Sarajedini \& Jablonka
2005, Worthey \etal 2005).  The standard solution is to allow for an
nonzero initial abundance for the gas that produces the first generation of
stars or a prompt enrichment mechanism.  A simple shift of initial
abundance has been demonstrated to be a poor fit for M31 and other nearby
galaxies (Harris \& Harris 2000, Sarajedini \& Jablonka 2005); however, an
accreting box scenario (a relaxing of the infall constraint, Harris \&
Harris 2000, dashed line) has been more successful and also lends itself to
an analytic solution.  For comparison to these models, we have replotted
the inner MDF of NGC 5128, a nearby elliptical (Harris \& Harris 2000) in
Figure 4.  Neither of these two models are a particularly good match to the
data.

A more recent model is given by an infall scenario (Kodama \& Arimoto
1997).  This model has the advantage of linking the accretion rate to the
star formation rate, a more physically realistic scenario, and yielding a
constant gas mass.  This model is shown in Figure 4, but also suffers
from an over abundance of metal-poor stars as compared to observations.
Adjustments to lower the number of metal-poor stars only results in an
over production of metal-rich stars, although the general shape of the
infall MDF follows the trend of the data.  This is primarily due to the
instantaneous mixing assumption for these models which smooths the
production of metals over space and time allowing for more metal-poor
stars.  In reality, one would expect regions of high metallicity to form,
which would produce a fast enrichment sequence.

The G-dwarf problem led us to consider a modification to the infall or
accreting models in a completely artificial fashion to match the MDF in NGC
5128.  This modification, which we call the `push' model, is a simple
reduction of the low metallicity end of the infall model.  To preform this
reduction, we adopt an infall model's shape, a peak metallicity with a
sharp high metallicity cutoff and a long low metallicity tail.  This
metallicity distribution's peak [Fe/H] is adjusted to alter the total mean
metallicity.  Our push model artificially reduces the low metallicity end
of this distribution, simply by a linear reduction while keeping the total
normalization constant (i.e. we push down the number of low $Z$ stars per
mass bin).  For our experiments herein, adequate fits to the data were
obtained with less than a 30\% reduction of the low metallicity side.

There is no direct theoretical or model support for our push model
(although it forces a relaxation of the instantaneous mixing assumption),
it is simply done to explore the effect of fewer low metallicity stars on
the integrated colors.  However, we note that the shape of the push model
reproduces the MDFs produced by inhomogeneous enrichment models (Malinie
\etal 1993, Oey 2000), where star formation occurs in discrete patches
throughout a galaxy and only allowed to mix between star formation
episodes.  This increases the amount of mixing and results in fewer
metal-poor stars.  As we can see in Figure 4, this type of model (with a
30\% reduction of metal-poor stars) produces the best `french curve'
through the NGC 5128 data.  Although our push model lacks a physical
foundation, we note that the inhomogeneous models also lack any simple
parameterization that relates to known galaxy properties and has a large
number of unconstrained variables.

\section{[Fe/H] Determination from the $<$Fe$>$ Index}

As stated in the Introduction, the power to line indice measurements (e.g.
the Lick/IDS system) is their direct determination of an element's
abundance.  One of the clearest measures of global metallicity, as outlined
in Trager \etal (2000), is the $<$Fe$>$ index, the numerical average of the
Fe5270 and Fe5335 lines.  This feature measures Fe, C, Mg, Ti, Si and,
thus, will be mildly sensitive to changes in the ratio of $\alpha$ elements
to Fe.  However, for the samples we have chosen (see below) the variation
in $\alpha$/Fe is small and will be ignored.

The $<$Fe$>$ index has been used by numerous studies of metallicity and age
in galaxies, but we have selected our sample, for comparison to our
narrowband colors, from three studies of early-type galaxies; Trager \etal
(2000), Poggianti \etal (2001) and Thomas \etal (2005).  Our choice of
these datasets is for a number of practical reasons.  One, the Trager \etal
work was a clear and strong step forward in the use of the Lick/IDS system
for galaxy work and sets the standard for age and metallicity determination
in that spectral system.  The Poggianti \etal study was on Coma cluster
galaxies where we have matching narrowband and near-IR colors (Rakos \&
Schombert 2008).  The Thomas \etal study is one of the most recent, and
largest, work with published $<$Fe$>$ values.  The total sample contains
185 galaxies of which we have matching color data for 119 galaxies.  For
the remaining 72 galaxies, we have estimated their $vz-yz$ colors using
their absolute $M_{5500}$ luminosity and the color-magnitude relation
(Odell, Rakos \& Schombert 2002), although their exclusion from the
analysis does not change our results.

The resulting plot of log $<$Fe$>$ versus metallicity color ($vz-yz$) is
shown in Figure 5.  Also plotted are the BC03 SSP models for a ages of 4
and 12 Gyrs.  What is immediately obvious from this plot is that a majority
of the metallicity data lies above and/or to the left of the SSP models.
As the SSP models accurately match the globular cluster color and
metallicities, then we might at first interpret the difference in Figure 5
as due to the composite nature of elliptical stellar populations.  However,
this diagram implies, logically, only that ellipticals are either 1) bluer
in $vz-yz$ color per metallicity value or 2) more metal-rich per integrated
color bin.  And these regions of the color-metallicity diagram could be
occupied for various star formation reasons.

\begin{figure}
\centering
\includegraphics[scale=0.95]{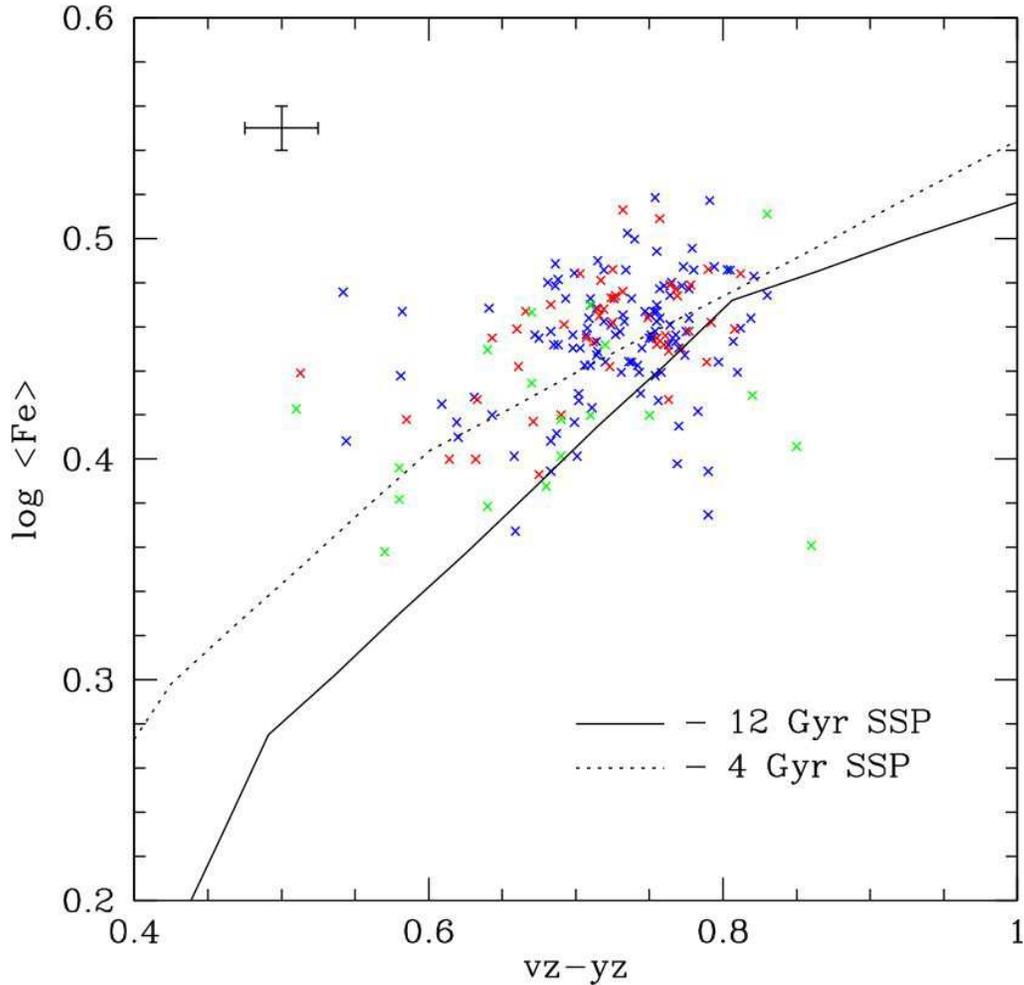}
\caption{The $<$Fe$>$ index from the Lick/IDS system versus galaxy
narrowband color, $vz-yz$.  The $<$Fe$>$ index is considered the most
accurate direct measure of the metallicity of a stellar population.  In
contrast, $vz-yz$ measures the mean metallicity color of a galaxy, a value
that is dominated by the effect of metallicity on the position of the RGB.
Blue symbols are data from Thomas \etal (2005), red symbols are from Trager
\etal (2000) and green symbols are Coma galaxies (Mobasher \etal 2001).
The solid line is a 12 Gyr SSP model from Bruzual \& Charlot (2003), the
dashed line is a 4 Gyr model.  The SSP models clearly do not fit the
metallicity data leading us to conclude that galaxies are either more
metal-rich for their color or too blue for their metallicity.  As
integrated color is more sensitive to the contribution of metal-poor stars
than the $<$Fe$>$ index, then a majority of the deviation from the SSP line
is due to color.  This diagram demonstrates that an SSP model is an
incomplete description of the colors of galaxies and a chemical evolution
scenario is required.
}
\end{figure}

For the first option, a bluer $vz-yz$ color can be derived at a constant
metallicity by a younger mean population, i.e. less than 4 Gyrs in age.  We
discuss the impact of younger stars in a later paper (Schombert \& Rakos
2009); however, to summarize that work, in order to match the color values
in Figure 5, the mean age of the entire underlying stellar population in a
majority of ellipticals would be required to be less than 2 Gyrs.  This is
extremely young for a total stellar population (although there may be small
numbers of young stars in ellipticals, see Trager \etal 2000) and is not
apparent in any other color system or spectral indicator.  In addition, it
would imply that a majority of ellipticals in intermediate redshift
clusters (0.3 $< z <$ 0.7) have star-forming colors, which is clearly not
seen (Rakos \& Schombert 1995).

The other logical option is that the metallicity indicator $<$Fe$>$ is
simply measuring a different quantity in galaxies than the integrated color
$vz-yz$, in this case a higher metallicity than indicated by the integrated
color.  This might be true due to the geometry of the data sampling, as
spectral values are based on core luminosities whereas the metallicity
color, $vz-yz$, is based on the total galaxy light.  Strong metallicity
gradients would result in a noticeable difference for metallicity values
determined by core light versus halo light.  In addition, composite
populations of varying metallicity may sum up in differing ways for
$<$Fe$>$ versus color (i.e. in particular the contribution from hot BHB
stars or a metal-poor MS turnoff population, see Worthey, Dorman \& Jones
1996).  We will explore both these effects in the next two sections.

\section{Aperture Correction to $<$Fe$>$}

A key difference between line indice studies and color work is
observational in that line indices, using the Lick/IDS system, are
determined by the smaller angular sized slit or fiber spectroscopy.  The
typical slit sizes are such that a line indice measurement of a particular
galaxy is going to be confined to the central regions.  Thus, due to the
geometry of the observational technique and the fact that spectroscopic
data is surface brightness weighted, the galaxy light obtained by line
indices studies will be heavily weighted towards the core regions.  Since
early-type galaxies have clear color gradients (Sanchez-Blazquez \etal
2006), which are known to be primarily due to metallicity gradients, this
leads to the possibility that line indice values are biased towards higher
metallicity values.  Thus, the comparison of line indice values to global
colors, those determined by the average metallicity as given by the entire
luminosity of the galaxy, may be invalid.

While it seems obvious that some bias towards higher metallicity values
exists in line indice studies, the amount of bias is unknown and may be
negligible.  Certainly, as colors gradients are known to be small in
early-type galaxies (Sandage \& Visvanathan 1978, Peletier \etal 1990,
Schombert \etal 1993), there is an expectation that with sufficient areal
coverage (e.g. over 1/3 an effective radius) the difference between a
global metallicity value and one determined from spectroscopy will be
small.

In order to estimate the metallicity bias for spectroscopic work due to an
aperture correction, we consider the Coma observations of Poggianti \etal
(2001) described in Mobasher \etal (2001).  This data was taken with a
fiber spectrograph with 2.7 arcsec slits (for comparison, SDSS uses three
arcsec fiber diameters).  At the redshift of Coma, this corresponds to a
diameter of 1.2 kpc.  For an $L_*$ galaxy, the light measured through this
aperture corresponds to approximately 1/3 the total light of the galaxy.
This will be less for brighter galaxies (larger effective
radii) and more for lower luminosity galaxies resulting in a variation of
about 20\% for the luminosity range given by the Poggianti \etal sample.

With the existence of color gradients, this smaller fraction of total light
measured by fiber slits will also contain a redder (more metal-rich)
stellar population.  Using the color gradients for our narrowband color,
$vz-yz$ (Schombert \etal 1993), we find that gradients take on a range of
values.  Galaxies with strong gradients (e.g. NGC 4374) have values of
${\Delta(vz-yz)}/{\Delta({\rm log} r)} = -0.15$.  Galaxies with weak
gradients (e.g. NGC 7562) have values near $-$0.05.  This results in
differences for mean color between the core luminosity seen by slits and
total color as $+$0.09 for strong gradients to $+$0.03 for weak gradients.
Converting this color difference into [Fe/H] (using BC03 12 Gyr SSP's)
leads to [Fe/H] values being 0.25 to 0.09 dex higher for spectroscopic
measurements compared to values deduced from a galaxy's total light.  Thus,
on average, the Lick/IDS values need to be lowered by approximately 0.15
dex to represent the mean [Fe/H] of a galaxy as a whole, which corresponds
to a change of 0.04 in the log $<$Fe$>$ index.

\section{Multi-Metallicity Population Correction to [Fe/H]}

A second correction to consider is that the value measured by colors is, of
course, a luminosity weighted integrated value.  Comparison to models has
always assumed that the underlying stellar population is simple (i.e. SSP).
There is every expectation that this is false based on any chemical
enrichment models which predicts a spread in metallicity.  In addition, HST
imaging has demonstrated broad MDF's in nearby galaxies (Worthey \etal
2005).

Again, using the SED models from Bruzual \& Charlot (2003) combined with an
infall scenario of chemical enrichment (Kodama as described in Yi \etal
1998), we can estimate the difference between the numerical averaged [Fe/H]
(actual sum of the metallicities of the stars) versus the luminosity
weighted value, $<$Fe/H$>$.  A series of simulations were run using this
formula where the only variable in this simulation is the peak [Fe/H] which
is allows to vary from $-$2.5 to $+$0.5.  The shape of the MDF is fixed,
starting at [Fe/H] = $-$2.5 and linearly adjusted to the peak [Fe/H].  This
MDF is then convolved with the SED models to produce colors for a composite
stellar population (CSP).  The output values from the CSP model are the
luminosity weighted $<$Fe/H$>$, what one would measure from the integrated
light and the actual numerical average metallicity, [Fe/H], from the sum of
the stars by mass.  For an SSP, or stellar population with a very narrow
range of metallicities, these values would be equivalent.  But for a
stellar population with a wide range of metallicities (in particular, a
long low metallicity tail), each metallicity bin contributes a slightly
difference luminosity per mass (higher for lower [Fe/H] values) and, thus,
the resulting observed $<$Fe/H$>$ value does not match the actual
underlying metallicity of the population by mass.  Since a metal poor
population is more luminous than the metal-rich population by mass, then a
luminosity weighted value of [Fe/H] will underestimate the true value
(Arimoto \& Yoshii 1987).

A series of conclusions were reached from comparing the values of a
luminosity weighted metallicity and the actual metallicity by stellar
number.  First, the exact shape of the MDF has little effect on the
correlation between average [Fe/H] and the luminosity weighted $<$Fe/H$>$
as long as there is a low metallicity component to the model.  Second, a
metal-poor tail is a requirement to the model as any distribution without a
metal-poor component failed to match the galaxy colors (e.g. the
color-magnitude relation).  Third, the relationship between the actual mean
[Fe/H] and the luminosity weighted (i.e. observed) metallicity is linear
and easy to calculate.  A correction can be defined between an observed
[Fe/H] value ($<$Fe/H$>$) and the true value which is expressed as:

$$ [{\rm Fe/H}] = 1.063 <{\rm Fe/H}> + 0.099 $$

Not to surprisingly, the inclusion of a metal poor tail to a metallicity
distribution causes the observed colors converted into a [Fe/H] value
(usually calculated from SSP models) to underestimate the real numerical
averaged [Fe/H].  Thus, for solar metallicities, the observed [Fe/H] must
be adjusted upward by approximately 0.1 dex.  Interestingly, this upward
correction to [Fe/H] is almost exactly balanced by a downward correction
needed for aperture corrections to line indice work (see previous section)
which would explain the high consistence between various studies.

Lastly, we can ask of the simulations the typical difference a metal-poor
component makes on the observed values of the $<$Fe$>$ index versus colors.
In the examples above, the typical change in color was on the order of 25\%
for a CSP model versus a SSP.  On the other hand, the corresponding change
in $<$Fe$>$ was only 10\%.  What this implies is that observed colors are
more strongly influenced by the hot component of a metal-poor population
(BHB stars), whereas the $<$Fe$>$ index derives most of its luminosity from
the RGB stars.

To summarize, the metallicity line indice, $<$Fe$>$, will suffer from both
an aperture effect, due to galaxy metallicity gradients, and an over
estimate of mean metallicity due to the use of SSP models with only a
single metallicity.  However, these two effects balance one another such
that the $<$Fe$>$ index is a good measure of the mean [Fe/H] of a galaxy's
stellar population.  Colors, on the other hand, being a total measure of
the integrated light of a galaxy, require a correction to estimate the mean
[Fe/H] since they will display the overall color of the composite stellar
population.  The correction is simple, dependent only weakly on the assumed
chemical enrichment model.  These effects also allow for an opportunity to
measure the effect of a composite population by comparing $<$Fe$>$ values
with a galaxies integrated color.

\section{Chemical Enrichment Interpretation of Color versus [Fe/H]}

Metallicity determination by colors requires an intermediary step, either
calibration by comparison to SED models or comparison to a standard system
such as galactic globulars.  For our continuum color system, the $vz-yz$
color is the metallicity indicator of choice since it has a linear
relationship with [Fe/H] that varies only slightly with age, as long as the
stellar population is older than 5 Gyrs (Rakos \& Schombert 2008).
For some subset of all three $<$Fe$>$ samples we have matching narrowband
photometry of the same galaxies for direct comparison.  These objects are
shown as data points in the following Figures, where the Thomas \etal
sample is shown as blue points, Trager \etal as red and Poggianti \etal as
green.

In order to test a CSP model of chemical enrichment, we have collected, as
discussed in \S3, a direct measure of [M/H] of the underlying stellar
population through the $<$Fe$>$ index and color for the underlying stellar
population as produced, primarily, by the mean [M/H] in our $vz-yz$
narrowband photometry.  As discussed in \S4 and 5, there are corrections to
be made to spectroscopic values to account for aperture effects and a
metal-poor component.  However, since the aperture corrections were matched
by opposing luminosity corrections (see \S5), no change was made to
$<$Fe$>$ data, and $<$Fe$>$ was converted to [M/H] using the prescriptions
outlined in Trager \etal (2000).  We will adopt these values as the mean
metallicity of the entire galaxy, a numerical average.  The model tracks
for the various chemical enrichment scenarios will use a mean metallicity
versus a luminosity weighted color, as the integrated colors reflect the
entire underlying stellar population.

\begin{figure}
\centering
\includegraphics[scale=0.75]{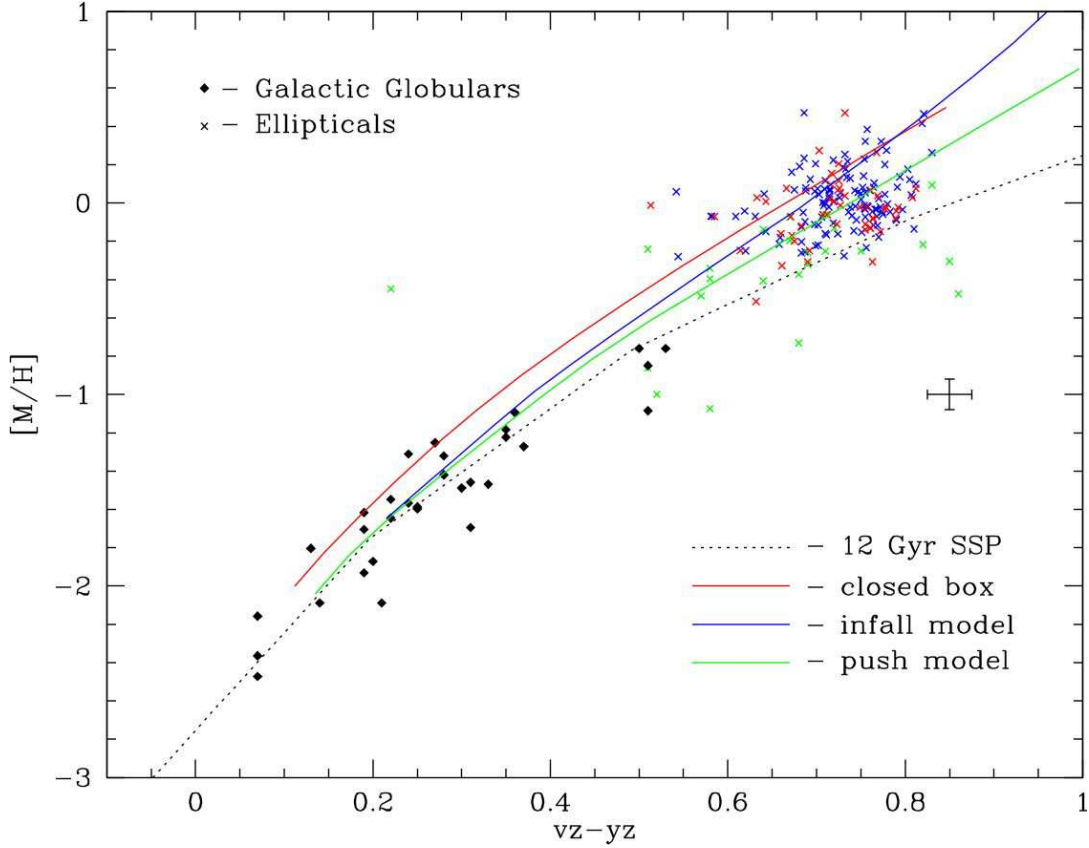}
\caption{The mean metallicity of a galaxy, [M/H], versus metallicity color,
$vz-yz$.  Galactic globulars are solid symbols, galaxies are crosses using
the same color scheme as Figure 5.  A black dashed line represents a 12 Gyr
SSP model, which is an excellent fit to the galactic globulars, but fails
for ellipticals.  The closed box and infall models are shown as solid red
and blue lines.  The closed box model fails to match the data, the infamous
G-dwarf problem, however, even the infall model appears to overproduce
metal-poor stars.  The best fit is found with our 'push' model, and
artificial suppression of the metal-poor tail to match the shape of the MDF
produced by inhomogeneous models (Oey 2000).
}
\end{figure}

The resulting plot of metallicity ([M/H] versus galaxy color ($vz-yz$) is
shown in Figure 6.  The color-metallicity relation (i.e. mass-metallicity
relation) is evident even though the mass range of the sample is limited.
Also plotted are the data for Galactic globular clusters (Rakos \&
Schombert 2005).  The SSP models for a 12 Gyr population of various
metallicities are shown as a dashed line.  While these models are excellent
fits to the GC data, they fail to describe the galaxy data.  A vast
majority of the galaxy data lie to the blueward side of the SSP models
indicating, again, that galaxies must be composed of a significant, at
least in luminosity, population of metal-poor stars.

The first chemical enrichment model to test is the closed box model as
outlined in Sarajedini \& Jablonka (2005) and shown as the red track in
Figure 6.  Of course, the immediate result of an enrichment model is the
addition of low metallicity stars to the integrated stellar population.
Using our CSP technique of summing a mixed stellar population, we find, not
surprisingly, that the integrated $vz-yz$ colors are bluer for a closed box
scenario than for a single metallicity SSP.  However, the closed box track
in Figure 6 is clearly too blue compared to the data on ellipticals.  This
is the famous G-dwarf problem (Pagel \& Patchett 1975), the known
deficiency of low metallicity stars in the solar neighborhood.  This
deficiency has also been noted in Milky Way halo populations (Tantalo \etal
1996), populations in M31 (Worthey \etal 2005) and NGC 5128 (Harris \&
Harris 2000).  The usual resolution is to modify the closed box assumption
with a model that has an initial enrichment component and an inflow of
metal-poor gas, an infall model.

The infall model assumes that gas flows into a system while the epoch of
star formation is still ongoing.  Thus, the gas is replenished at the same
rate as star formation consumes it (Gibson \& Matteucci 1997).  For our
purposes, we adopted the Kodama infall model outlined in Yi \etal (1998).  We
parameterized the models over galaxy mass by sliding the metallicity
distribution shape over the metallicity range [M/H] = -2 to $+$1 as
described in Rakos \etal (2001). The resulting CSP models using an
infall scenario of varying total mass is shown as the blue line in Figure
6.  While the infall model reduces the number of metal-poor stars, and
thereby reddening the predicted $vz-yz$ colors, this effect does not match
the elliptical data.

The failure of the infall model, in that it still appears to produce too
many metal-poor stars per gas mass, motivated us to produce a model which
suppresses the low metallicity tail.  We refer to this model as the 'push'
model as it pushes down the low end of the metallicity curve (see \S2.3).
However, this is a completely artificial change to the infall model, and
has no physical basis other than it results in the color changes needed to
match the data.   Our push model, shown in Figure 6 and the resulting
color-metallicity track is shown in green in Figure 6, is a good match for
the data with a 30\% reduction to the metal-poor component.  This agrees
well with our push model fit to the NGC 5128 data, a similar morphology
type to our elliptical sample.  This also confirms that the G-dwarf problem
is even more severe in ellipticals than spirals such as the Milky Way and
M31 (Worthey, Dorman \& Jones 1996).

A sharp reduction of the low metallicity end of a galaxy's MDF is a feature
to inhomogeneous enrichment models.  These models relax the chemical
homogeneity assumption by adopting a fixed dispersion in metal production.
Following the paradigm of Oey (2000), these models are parameterized by two
variables 1) the number of generations of star formation and 2) the filling
factor in the ISM that each generation occupies.  As noted in the Oey
study, an old, metal-rich population is achieved by a high filling factor.
To reproduce our push model values would require a low number of star
formation generations, i.e. a rapid and short initial star formation epoch
for ellipticals in agreement with the conclusions based on $\alpha$/Fe
ratios and galaxy mean ages (Rakos, Schombert \& Odell 2008).

\section{Conclusions}

Combining information from spectroscopic measurements of galaxy cores with
narrowband colors allows for a test of two, relatively independent,
estimators of mean metallicity.  For example, spectroscopic lines measure
key metallicity lines (e.g. Fe) directly, whereas, colors measure the
effect of changing metallicity on temperature of the stellar population's
atmospheres (primarily the position of the RGB in an HR diagram).  This
different sensitivity to metallicity can be exploited to test predicts from
chemical enrichment models on the shape of the MDF.

We summarize our results as follows:

\begin{itemize}

\item We have examined the accuracy of SED models on predicting narrowband
and near-IR colors in globular clusters, simple stellar populations of
singular age and metallicity.  We find that our optical colors and $V-K$
are well matched by SED models; however, near-IR colors (i.e. $J-K$) do not
follow SSP tracks.

\item The most accurate measure of metallicity (e.g. [Fe/H]) is our
narrowband $vz-yz$ color.  Based on arguments in Schombert \& Rakos (2009),
the range of galaxy ages from 8 to 13 Gyrs has a negligible effect on the
[Fe/H] versus $vz-yz$ correlation.  In addition, Schombert \& Rakos (2009)
rejects the proposal that a majority of galaxy ages in clusters are of less
than 8 Gyrs.  Thus, for this limited set of galaxy morphology (early-type),
we can use the $vz-yz$ color as a sole measure of integrated [Fe/H] in a
composite stellar population.

\item Comparison of optical and near-IR elliptical colors demonstrates that
early-type galaxies are best explained by a composite stellar population.
As we reject extremely young mean galaxy ages, then, in order to match the
colors of ellipticals, the underlying stellar population must be singular in
age but with a range of internal metallicities.

\item There is a well defined continuum between $vz-yz$ color and
metallicity ([M/H]) where the colors of ellipticals are bluer than that
predicted by SSP models.  The inclusion of a metal-poor stellar population
is an obvious solution, where the fraction of metal-poor stars can be
estimated from a chemical evolution scenario.

\item The simplest chemical evolutionary scenarios, the closed box model
and the initial enrichment model can be rejected as solutions to the MDF in
ellipticals due to their over production of metal-poor stars (and resulting
blue integrated colors).  This results in the infamous G-dwarf problem
(Pagel \& Patchett 1975), a well known problem for the local stellar
neighborhood and stellar populations in nearby galaxies.  Our data
indicates that the G-dwarf problem is universal (Worthey, Dorman \& Jones
1996).

\item Infall models, while a better match to the data, also over produce
metal-poor stars and lie on the blue side of the data.  Our analytic 'push'
model is a artificially constructed curve that suppresses the number of
metal-poor stars to the typical infall model.  This MDF shape, narrower at
its peak, reduced on the metal-poor end and sharper on the metal-rich end,
matches the elliptical galaxy data.  These types of curves are also
predicted by inhomogeneous models of chemical evolution (Tinsley 1975,
Malinie \etal 1993, Oey 2000) with high filling factors and rapid initial
epochs of star formation.

\end{itemize}

A key missing piece of the chemical evolution puzzle in ellipticals is the
age-metallicity relationship (AMR).  The AMR would provide a detailed
breakdown of the evolution of the MDF, however, this information will be
difficult to extract from composite systems such as distant ellipticals.
Another avenue for exploration is the presence of metallicity gradients.
With guidance by a galaxy formation scenario (matching our chemical
enrichment scenarios), one could, ideally, match the metallicity
distribution as a function of radius mapped into time.  Future work with
our narrowband system will concentrate on spatial analysis of nearby
ellipticals to this very end.

\acknowledgements

Financial support from Austrian Fonds zur Foerderung der Wissenschaftlichen
Forschung and NSF grant AST-0307508 is gratefully acknowledged.  We thank
all the various observatories which have supported our efforts, KPNO, CTIO
and ESO.  This research has made use of the NASA/IPAC Extragalactic
Database (NED) which is operated by the Jet Propulsion Laboratory,
California Institute of Technology, under contract with the National
Aeronautics and Space Administration and has made use of data obtained from
or software provided by the US National Virtual Observatory, which is
sponsored by the National Science Foundation..

\end{document}